\documentclass{PoS}

\title{A new MicroTCA-based waveform digitizer for the Muon g-2 experiment}

\ShortTitle{A new MicroTCA-based waveform digitizer for the Muon g-2 experiment}

\author{
\speaker{David A. Sweigart}\thanks{For the Muon g-2 collaboration.}\\
Cornell University\\
E-mail: \email{das556@cornell.edu}
}

\abstract{We present the design of a new $\mu$TCA-based waveform digitizer, which will be deployed in the Muon g-2 experiment at Fermilab and will allow our pileup identification requirement to be met.  This digitizer features five independent channels, each with 12-bit, 800-MSPS digitization and a 1-Gbit memory buffer.  The data storage and readout along with configuration are handled by six Xilinx Kintex-7 FPGAs.  In addition, the digitizer is equipped with a mezzanine card for analog signal conditioning prior to digitization, further widening its range of possible applications.  The performance results of this design are also presented, highlighting its $0.51 \pm 0.13$ mV intrinsic noise level and $< 22$ ps intrinsic timing resolution between channels.  We believe that its performance, together with its flexible design, could be of interest to future experiments in search of a cost-effective waveform digitizer.}

\FullConference{
38th International Conference on High Energy Physics\\
3-10 August 2016\\
Chicago, USA
}

\begin{document}

\section{Muon g-2 pileup requirement}

The Muon g-2 experiment at Fermilab aims to measure the muon's anomalous magnetic moment, $a_\mu$, to 140~ppb --- a four-fold improvement over the latest measurement, BNL E821, which was statistics-limited \cite{e821}.  In such experiments, a polarized beam of positive muons is stored inside a ring with a uniform magnetic field.  The decay positrons are then detected by 24 electromagnetic calorimeters stationed around the storage ring.  In the laboratory frame, the number of high-energy positrons oscillates at the cyclotron minus the spin precession frequency, thus allowing us to extract $a_\mu$ by fitting the number of detected positrons in time above an optimized energy threshold \cite{tdr}.

The $a_\mu$ precision goal will be achieved, in part, by quadrupling the number of stored muons for each measurement period.  This would cause a 16-fold increase in pileup --- events with multiple positrons that are indistinguishable from one higher-energy positron.  The time-dependence of the pileup fraction poses a direct bias to our $a_\mu$ measurement, which scales approximately linearly with the positron rate on each detector element \cite{tdr}.  The Muon g-2 experiment strives to reduce the intrinsic pileup fraction nine-fold.  To that end, we have designed a custom waveform digitizer to separate pileup in time three-fold over that used in the previous experiment, by increasing the bit depth from 8 to 12 bits and the sampling rate from 400 to 800 MSPS.  Monte Carlo models predict our lowest pileup temporal separation to improve from 5 to $\le 3.5$ ns, with the remaining pileup reduction coming from spatial separation with new segmented PbF$_2$ calorimeters \cite{tdr}.

\section{$\mu$TCA-based architecture}

The waveform digitization system was built around $\mu$TCA technology, with each waveform digitizer conforming to the Advanced Mezzanine Card base specification.  This industry standard allows the digitizer to be installed into an off-the-shelf $\mu$TCA chassis, which provides a common platform for interconnections, cooling, and power and clock distribution.  We specifically use an Aluminum VadaTech VT892 chassis, which can support up to 12 digitizers.  A VadaTech UTC002 $\mu$TCA Carrier Hub (MCH) is used to supply Ethernet and IPMI communication to each digitizer via the $\mu$TCA backplane for configuring and monitoring.  In place of a redundant MCH, we install an AMC13 module into each chassis, designed by Boston University for the CERN CMS experiment \cite{amc13}.  The AMC13 is responsible for distributing the experiment clock, triggers, and controls and for transmitting each digitizer's data to the data acquisition system.

\section{Waveform digitizer design}

The waveform digitizer design features five independent channels, as shown in Fig.~1.  Each digitization channel is centered on a 12-bit, 800-MSPS ADC (TI ADS5401) and is equipped with a 1-Git DDR3 SDRAM (Micron MT41J64M16), which can buffer over nine seconds of data.  The data acquisition logic for each channel is controlled by its own Kintex-7 FPGA (Xilinx 7K70T).  A sixth Kintex-7 FPGA (Xilinx 7K160T) communicates with each of these FPGAs over a dedicated 5-Gbit/s serial link to provide controls and to read out their buffered data sequentially, which are transferred via the $\mu$TCA backplane's 5-Gbit/s Fabric A link to the AMC13 with 8b/10b encoding.  This FPGA also interfaces to the MCH over a 10-Gbit/s Ethernet link using IPbus --- an IP-based protocol used for configuring and monitoring $\mu$TCA modules \cite{ipbus}.  Per $\mu$TCA standard, the MCH further communicates over IPMI with a micro-controller (Atmel AT32UC3A1) on each digitizer, which serves as its $\mu$TCA Management Controller (MMC).

\begin{figure}[t]
\centering
\includegraphics[width=\textwidth]{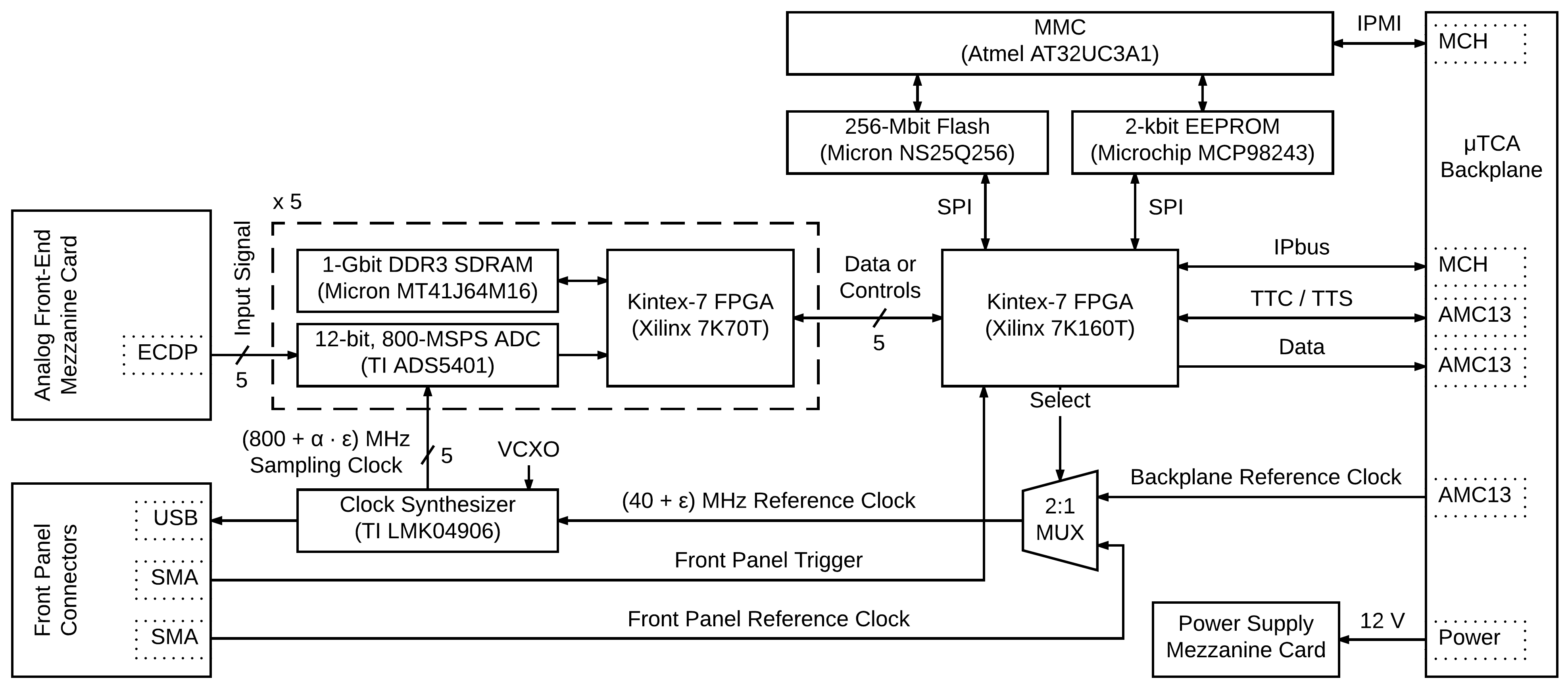}
\caption{The hardware architecture of the five-channel waveform digitizer for the Muon g-2 experiment.}
\label{fig:design}
\end{figure}

The AMC13 will receive the experiment's blinded $40+\epsilon$ MHz master clock via optical fiber using the CERN TTC protocol \cite{ttc}.  This reference clock is fanned out over the $\mu$TCA backplane's CLK1 link to each digitizer.  An on-board clock synthesizer (TI LMK04906), operating in its dual PLL mode, provides the frequency upconversion to the ADC sampling rate for each channel independently, along with a sixth output that is routed to the front panel for monitoring.  The nominal $800+\alpha \cdot \epsilon$ MHz rate, where $\alpha$ is the upconversion factor, will be used for all channels digitizing calorimeters, but the board design does allow for configurable rates as low as 40 MHz.  Each clock output can also be delayed in 25-ps steps, allowing for channel-to-channel timing corrections.

In addition to the reference clock, the CERN TTC protocol encodes the experiment controls and triggers, which are fanned out via the $\mu$TCA backplane's Fabric B link to each digitizer.  The controls, in part, allow for switches between operation modes, as described below; the triggers are used to initiate the buffering of ADC samples.  The trigger logic for each channel is independent, allowing for all or a subset of channels to be triggered.  For added flexibility, the reference clock as well as the triggers can be supplied through two SMA connectors on the front panel.

The waveform digitizer design also features two mezzanine cards.  The first card controls the power supply and regulation, allowing for an easy replacement upon failure and minimizing overall risk.  The second card is an analog front-end for input-signal conditioning, providing inexpensive customization of the digitizer.  In our design, the differential input signal is bandwidth limited to 230 MHz for noise reduction and is DC-coupled for a rate-independent pedestal.  Three daisy-chained digital-to-analog converters are also used to supply programmable offsets to the pedestal.

Furthermore, the FPGA firmware allows for flexible data acquisition.  In the main operation mode, when a TTC trigger is received, ADC data are buffered according to a regular pattern --- with the number of sampling windows, the gap between sampling windows, and the sampling window length specified in configuration registers.  Up to three different acquisition patterns can be stored; the specific pattern to use for subsequent triggers is specified by a TTC control command.  Alternatively, the digitizer can operate in an asynchronous mode, where the ADC data are stored continuously into a circular memory block in each channel's FPGA.  When a front-panel trigger is received, a preconfigured number of ADC samples before and after the trigger arrival are stored in the DDR3 SDRAM, which can then be read out when initiated by a TTC control command.

\section{Characterization results}

\begin{figure}[t]
\centering
\includegraphics[height=0.25\textwidth]{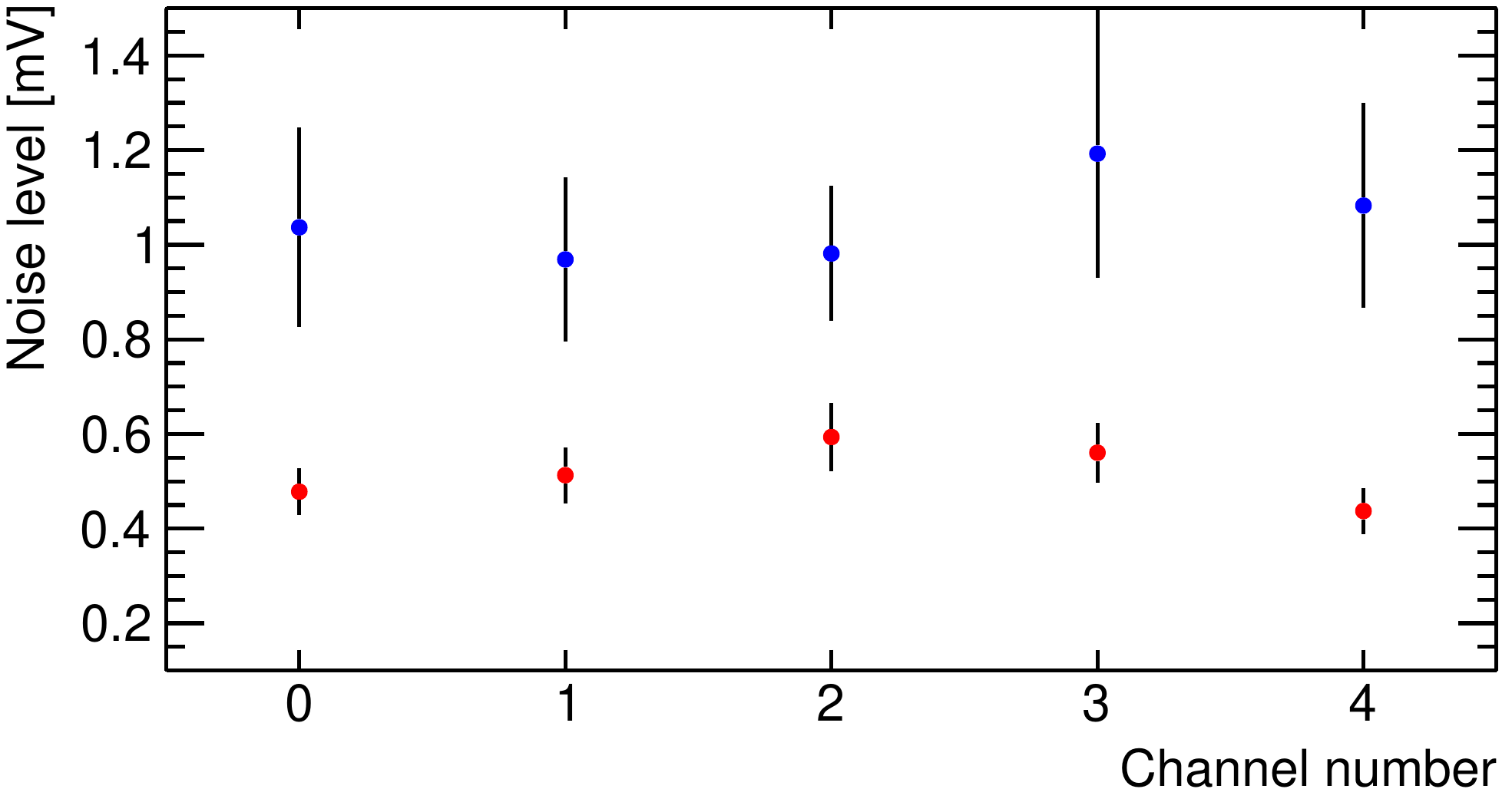}
\hfill
\includegraphics[height=0.25\textwidth]{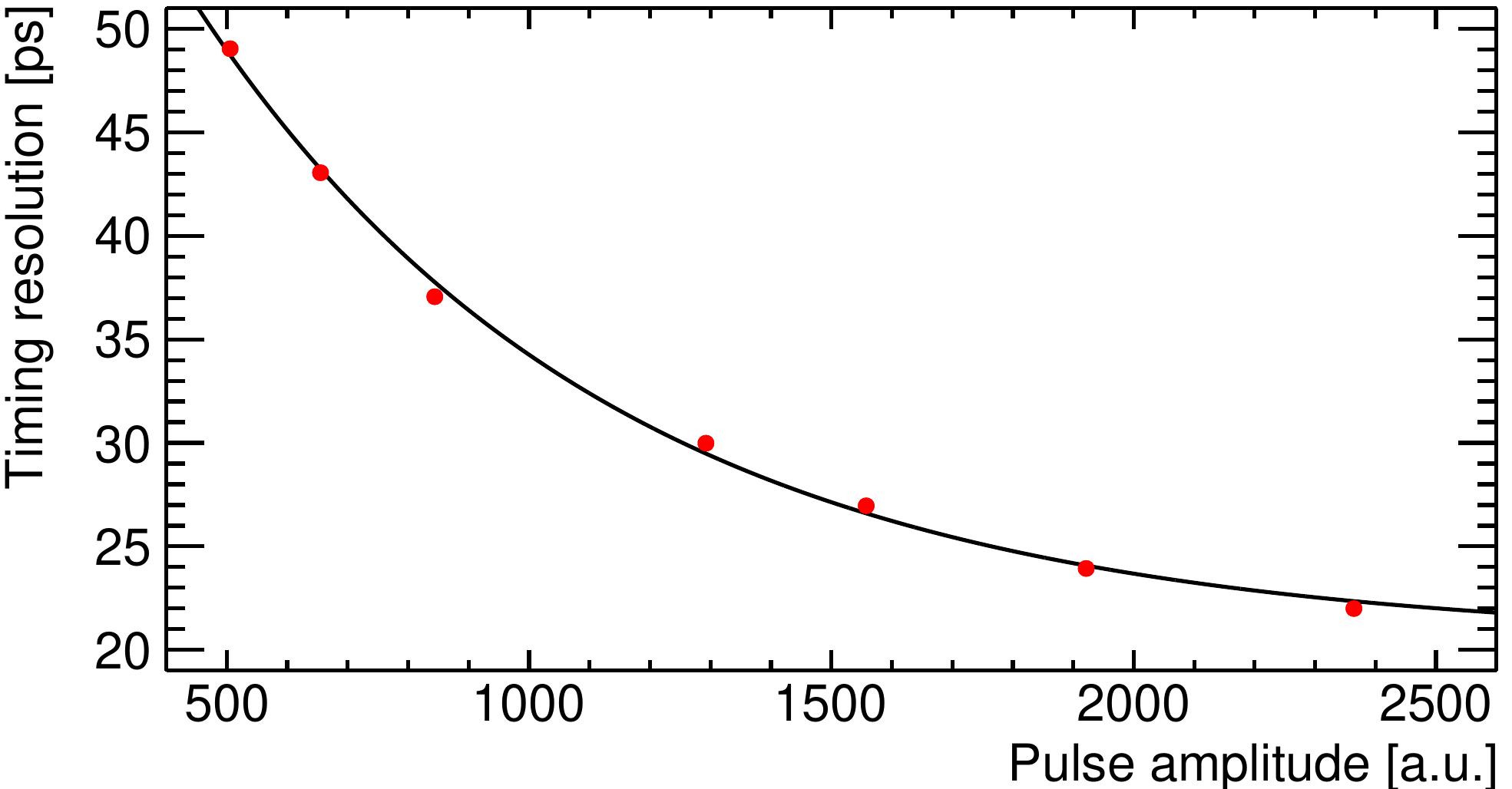}
\caption{Left: Intrinsic (red) and full system (blue) noise level, averaged for 277 and 11 digitizers, respectively.  Right: Timing resolution between channels for varying amplitudes; the fit serves only as a guide.}
\label{fig:results}
\end{figure}

We have produced 323 five-channel waveform digitizers passing initial quality control, which will be deployed in the Muon g-2 experiment at Fermilab.  In July 2016, the full calorimeter system was evaluated using the electron beam at the End Station Test Beam at SLAC, stress-testing 18 of these digitizers extensively with no faults.  As shown in Fig. 2, the intrinsic board noise level --- as measured when the signal input is shorted --- is $0.51 \pm 0.13$ mV among channels.  The full system noise level was found to be $1.05 \pm 0.46$ mV among channels, which allows us to achieve $\le5$ \% energy resolution at 2 GeV for the calorimeter.  The timing resolution is also shown in Fig. 2 to be $<50$ ps --- 4 \% of an 800-MHz clock tick --- for reasonable pulse amplitudes.  The upconversion scale factor, $\alpha$, has separately been measured to 0.1 ppb --- an order of magnitude lower than the experiment requires --- using multiple evaluation boards.  In addition, the digitized pulse is found to scale linearly with the input pulse, having an $\mathcal{O}(10^{-5})$ quadratic term for each board.  We have also observed neither crosstalk between channels nor performance differences within a 46-G field.

\paragraph{Acknowledgements}

This research was supported, in part, by the U.S. National Science Foundation's MRI program PHY-1337542 and by the U.S. Department of Energy Office of Science, Office of Nuclear Physics under award number DE-FG02-97ER41020.


\begin{thebibliography}{5}

\bibitem{e821}
G. Bennett \textit{et al.} (E821 Collaboration),
{Phys. Rev. D} \textbf{73}, 72003 (2006),
[\texttt{hep-ex/0602035}].

\bibitem{tdr}
J. Grange \textit{et al.} (E989 Collaboration),
\textit{Muon g-2 Technical Design Report}
(2015),
[\texttt{1501.06858}].

\bibitem{amc13}
E. Hazen \textit{et al.},
{J. Instrum.} \textbf{8}, C12036 (2013).

\bibitem{ipbus}
C. Ghabrous Larrea \textit{et al.},
{J. Instrum.} \textbf{10}, C02019 (2015).

\bibitem{ttc}
B. G. Taylor,
{IEEE Trans. Nucl. Sci.} \textbf{45} (3), 821 (1998).

\end{thebibliography}
\end{document}